\documentclass[aps,prd,reprint,superscriptaddress]{revtex4-1}
\usepackage{graphicx}
\usepackage{dcolumn}
\usepackage{natbib,aas_macros}
\usepackage{color}

\begin{document}
\title{Ultrahigh-energy Cosmic-ray Nuclei from Black Hole Jets:\\
Recycling Galactic Cosmic Rays through Shear Acceleration}

\author{Shigeo S. Kimura}
\affiliation{Department of Physics, The Pennsylvania State University, University Park, Pennsylvania 16802, USA}
\affiliation{Department of Astronomy \& Astrophysics, The Pennsylvania State University, University Park, Pennsylvania 16802, USA}
\affiliation{Center for Particle and Gravitational Astrophysics, The Pennsylvania State University, University Park, Pennsylvania 16802, USA}
\author{Kohta Murase}
\affiliation{Department of Physics, The Pennsylvania State University, University Park, Pennsylvania 16802, USA}
\affiliation{Department of Astronomy \& Astrophysics, The Pennsylvania State University, University Park, Pennsylvania 16802, USA}
\affiliation{Center for Particle and Gravitational Astrophysics, The Pennsylvania State University, University Park, Pennsylvania 16802, USA}
\affiliation{Yukawa Institute for Theoretical Physics, Kyoto, Kyoto 606-8502 Japan}
\author{B. Theodore Zhang}
\affiliation{Department of Astronomy; School of Physics, Peking University, Beijing 100871, China}
\affiliation{Kavli Institute for Astronomy and Astrophysics, Peking University, Beijing 100871, China}
\affiliation{Department of Physics, The Pennsylvania State University, University Park, Pennsylvania 16802, USA}

\date{\today}

\begin{abstract}
We perform Monte Carlo simulations of transrelativistic shear acceleration dedicated to a jet-cocoon system of active galactic nuclei. A certain fraction of galactic cosmic rays in a halo is entrained, and sufficiently high-energy particles can be injected to the reacceleration process and further accelerated up to 100 EeV. We show that the shear reacceleration mechanism leads to a hard spectrum of escaping cosmic rays, $dL_E/dE\propto E^{-1}-E^0$, distinct from a conventional $E^{-2}$ spectrum. The supersolar abundance of ultrahigh-energy nuclei is achieved due to injections at TeV-PeV energies. As a result, we find that the highest-energy spectrum and mass composition can be reasonably explained by our model without contradictions with the anisotropy data. 
\end{abstract}

\maketitle

\section{Introduction}

The origin of ultrahigh-energy cosmic rays (UHECRs) has been under intense debate for more than half a century~\cite{Linsley:1963km}. 
Observationally, remarkable developments was made by High-Resolution Fly's Eye, Pierre Auger Observatory (PAO), and Telescope Array (TA)~\cite{Kotera:2011cp}. 
The spectrum of UHECRs has a cutoff around 60 EeV~\cite{Abbasi:2007sv,Abraham:2008ru}, consistent with the energy of the Gresien-Zatspin-Kuzmin cutoff for protons~\cite{Greisen:1966jv,Zatsepin:1966jv} or a photodisintegration cutoff for irons. 
The absence of small-scale anisotropy in the arrival direction of the highest-energy cosmic rays (CRs) places a lower limit on the number density of UHECR sources, $n_s\gtrsim{10}^{-6}-{10}^{-5}{\rm~Mpc}^{-3}$~\cite{Kashti:2008bw,Takami:2008rv,Takami:2012uw,PierreAuger:2014yba}, which may be stronger at 10~EeV~\cite{Takami:2014zva}. 
The CR composition is estimated from the depth of the shower maximum, $X_{\rm{max}}$~\cite{Abraham:2010yv,Aab:2014kda,Aab:2015bza,Fukushima:2015bza}. The results of PAO and TA on the mean depth $\langle{X_{\rm{max}}}\rangle$ seem compatible~\cite{Abbasi:2015xga}. However, the interpretation of the data is controversial, partly due to uncertainty in hadron physics implemented in extensive air-shower simulations. 
The latest hadron interaction models imply that the composition gradually becomes heavier for $\gtrsim3$~EeV. On the other hand, UHECRs are believed to be dominated by light elements around $1-3$~EeV~\cite{Abbasi:2009nf,Abbasi:2014sfa}.

The heavy mass composition, if true, challenges astrophysical models for UHECR sources. 
The simultaneous fittings of the spectrum and composition suggest (i) a mass composition heavier than the solar abundance and (ii) a spectrum harder than a conventional $E^{-2}$ spectrum for the plausible redshift evolution~\cite{Allard:2011aa,Taylor:2015rla,Aab:2016zth}. The former difficulty is more serious if any anisotropy is established for the highest-energy CR nuclei, since the similar level of the anisotropy is expected at the same rigidity of protons ($\sim1-10$~EeV)~\cite{Lemoine:2009pw,Abreu:2011vm,Liu:2013ppa} (see also~\cite{Takami:2008rv}). 
Gamma-ray bursts \cite{Murase:2008mr,Horiuchi:2012by,Wang:2007xj,Globus:2014fka} and newborn pulsars~\cite{Fang:2012rx,Fang:2013vla} can provide a metal-rich composition, but UHECR nuclei must survive against photodisintegration in the sources and the intrinsic abundance ratio is essentially treated as a free parameter. 
Also, the luminosity argument~\cite{Blandford:1999hi} and the nondetection of ultrahigh-energy neutrinos~\cite{Aartsen:2016ngq} disfavor steady UHECR proton sources~\cite{Murase:2008sa,Takami:2014zva,Fang:2016ewe}. Steady sources, such as active galactic nuclei (AGNs), can accelerate CR nuclei up to 100~EeV (e.g.,~\cite{Murase:2011cy,Murase:2014foa}), but the origin of heavy composition with a hard spectrum has been unclear. 

\begin{figure}
\includegraphics[width=0.8\linewidth]{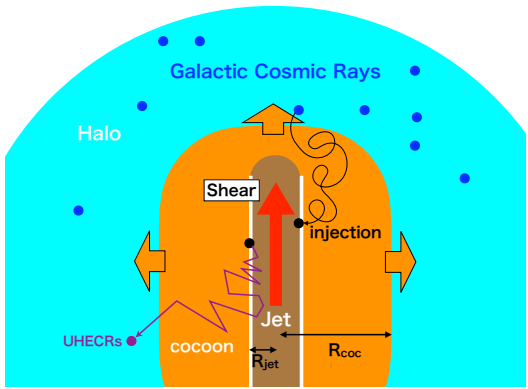}
\caption{The schematic picture of shear acceleration in a jet-cocoon system of an AGN. A fraction of GCRs swept up by the flow can be accelerated up to ultrahigh energies. 
\label{fig:schematic}}
 \end{figure}

In this work, we provide a new scenario that simultaneously explains the spectrum and composition, overcoming the above difficulties. 
First, in Section \ref{sec:shearaccel}, we consider the shear acceleration of CRs around transrelativistic shear layers, where both discrete and continuous shear acceleration mechanisms are discussed. 
For high-energy CRs, we perform detailed numerical simulations and  show that each composition species of the CRs leaving the accelerators have a hard spectrum with a rigidity dependent energy cutoff at $E_{i,\rm{max}}=Z_iE_{p,\rm{max}}$ (i.e. Peters cycle~\cite{Peters:1661aa,Gaisser:2013bla}). Here $E_{i,\rm{max}}$ is the ion maximum energy at the CR accelerators, $E_{p,\rm{max}}$ is the proton maximum energy, and $Z_i$ is the particle charge for a particle species with $i$. For low-energy CRs, we discuss the analytical CR spectrum, and show that the high-energy CRs can be accelerated mainly via the discrete shear acceleration mechanism.
Then, in Section \ref{sec:recycling}, we apply the mechanism to the system that is composed of an AGN jet and a cocoon inflated by the jet. We find that TeV-PeV CRs injected from a galactic halo are naturally accelerated by the shear acceleration, which can generate UHECRs with energies up to 100~EeV (Fig.~\ref{fig:schematic}). We also calculate the UHECR propagation in intergalactic space, and demonstrate that our model accounts for the observed Auger data well. In Section \ref{sec:summary}, we summarize our results and discuss implications.

\section{Shear Acceleration}\label{sec:shearaccel}

Shear acceleration is a class of Fermi acceleration mechanisms~\cite{Berezhko:1981aa,Earl:1988aa,Webb:1989aa,Ostrowski:1990aa,Rieger:2006uv}. The shear acceleration occurs when the relativistic particles are inside an ordered shear velocity field, which is commonly expected in the astrophysical jets~\cite{Ostrowski:1998ic,Rieger:2004jz,Rieger:2016ewr} and accretion flows~\cite{Katz:1991aa,Subramanian:1998hz,Kimura:2016fjx}. 
When the shear is continuous in the scale of the mean free path for scatterings with magnetic fields, the acceleration mechanism is basically the same as the stochastic acceleration in a turbulence. A particle that has a head-on (tail-on) collision gains (loses) energy, and the particles are statistically accelerated because the head-on collision is more probable than the tail-on collision~\cite{Fermi:1949ee,Subramanian:1998hz}.
When the scattering mean free path is longer than the scale of the shear velocity gradient, the acceleration is regarded as the Fermi process in the discrete shear~\cite{Ostrowski:1990aa,Ostrowski:1998ic}. In our scenario, UHECR production proceeds in this regime due to their large Larmor radii. The spatial diffusion is important, so that we take a numerical approach to properly consider the geometry.
Note that the continuous shear acceleration and discrete shear acceleration are different in terms of the properties of CR acceleration, which leads to the important difference in their time scales such as the CR escape time and CR acceleration time. This may result in distinct predictions for CR spectra.

\subsection{Discrete shear acceleration}

\subsubsection{Setup for Monte Carlo simulations}

We consider a jet-cocoon system~(see, e.g.,~\cite{Begelman:1989jp,Kawakatu:2006qc}).
To mimic the geometry of interest (see Fig.~\ref{fig:schematic}), we consider two cylinders with radii of $R_{\rm{jet}}$ and $R_{\rm{coc}}$. We parameterize the cocoon radius as $R_{\rm{coc}}\equiv\xi_{c/j}R_{\rm{jet}}$. The shear between the jet and cocoon is given by the jet velocity, $c\beta_{\rm{jet}}$. The cocoon is quasi-spherical in general. For simplicity, we assume the vertical length of the jet and the cocoon to be equal to the cocoon radius: $l_{\rm{jet}}=l_{\rm{coc}}=R_{\rm{coc}}$, which is sufficient for the purpose of this work.

We expect that both of the jet and cocoon have turbulent magnetic fields 
that scatter the particles.
We can parameterize the mean free path inside the cocoon as 
$\lambda_{i,\rm{coc}}=(E/E_{i,\rm{coh}})^{\delta}l_{\rm{coh}}$, where $l_{\rm{coh}}$ is the coherence length and $E_{i,\rm{coh}}=Z_ieB_{\rm{coc}}l_{\rm{coh}}$ ($B_{\rm{coc}}$ is the magnetic field strength in the cocoon). 
The particles are resonantly scattered by turbulence for $E<E_{i,\rm{coh}}$, which leads to $\delta=1/3$ if we assume the Kolmogorov turbulence inside the cocoon \cite[e.g.][]{Stawarz:2008sp}.
On the other hand, particles are scattered in non-resonant manner with small-scale turbulence for $E>E_{i,\rm{coh}}$, resulting in $\delta=2$ \cite[e.g.][]{Sironi:2013ri}.
Both the turbulence and magnetic field are likely to be strong in the jet, 
and the diffusion process in the strong turbulence is likely to be the Bohm limit \cite{Roh:2015cxa,Kimura:2016fjx}. 
Thus, we use the Bohm limit there, $\lambda_{i,\rm{jet}}=E/(Z_ieB_{\rm{jet}})$, where $B_{\rm{jet}}$ is the magnetic field strength in the jet. 
The particles move in a manner of the random walk by these interactions, and undergo multiple passage through the shear layer. This results in the discrete shear acceleration.  

For a given nuclear species, we inject 262,144 particles with an injection energy of $E_{i,\rm{inj}}$ (see Section \ref{sec:injection}) at the jet-cocoon boundary at $t=0$, and track them by a time of $t=t_{\rm{ad}}\approx{R_{\rm{coc}}}/v_{\rm{exp}}$, where $v_{\rm{exp}}$ is the expansion velocity of the cocoon. After this time scale, we expect that the particles lose their energies due to the adiabatic expansion. Since injected particles are reaccelerated to ultrahigh energies, more than 89 \% of the particles escape from the system by the end of simulation runs.
The number of the injected particles is normalized by the injection rate $\dot{N}_{\rm{inj}}$  (see Section \ref{sec:injection}). 
The particles travel straightly until they are scattered by a magnetic field. The scattering angle distribution is assumed to be isotropic in the rest frame of each fluid, which is a simplified but reasonable approximation in our problem, given that almost all the particles experience many scatterings during their residence time (cf.~\cite{Kato:2000hd,Kato:2003st,Aoi:2007aj} and references therein).  When the particles diffuse out beyond the cocoon radius, $R_{\rm{coc}}$, or the jet length, $l_{\rm{jet}}$, they are recorded as ``escaping'' particles. 

Hereafter, we consider radio-loud AGNs, in particular Fanaroff-Riley I radio galaxies (FR Is), to demonstrate our results (see Appendix \ref{sec:radio-quiet} for an application to radio-quiet AGNs). Powerful kiloparsec-scale jets are commonly seen in radio galaxies, and they are often accompanied by radio lobes or bubbles. The jets sweep up the circumgalactic materials in galactic halos, and eventually propagate into the intergalactic medium. The plasma inflated by the jet forms a cocoon, which is attributed to a radio lobe or bubble. The length of the jet depends on the age of AGN, and we consider the time when the jet finishes sweeping the halo in which galactic CRs (GCRs) are confined, i.e., $l_{\rm{jet}}=H_h$, where the scale height of the CR halo $H_h$ is set to 5~kpc~\cite{Strong:2007nh}. Kiloparsec-scale jets of FR Is are only mildly relativistic, so that the jet velocity is set to $\beta_{\rm{jet}}=0.7$~(e.g.,~\cite{Bowman1996,Laing:2002aa,Canvin2005:aa}). 
The ratio of the cocoon to jet radii is given by $\xi_{c/j}=10$ as a reference value~\cite{Mizuta:2004gu,Rossi:2008cm}, which leads to $R_{\rm{jet}}=0.5$~kpc. The magnetic fields are assumed to be $B_{\rm{jet}}=0.3\rm~mG$ (e.g.,~\cite{Stawarz:2005tq}) and $B_{\rm{coc}}=3~\mu$G~\cite{Dunn:2005xk,Kataoka:2004at}. The expansion velocity of the cocoon is set to $v_{\rm{exp}}=3000\rm~km~s^{-1}$~\cite{Bordas2011:aa}. For the coherence length, we use $l_{\rm{coh}}=0.03R_{\rm{coc}}$ as a reference value. These fiducial parameters are consistent with the observations of radio galaxies, and the spectral shape is largely unaffected by the change of the parameters.

\subsubsection{Maximum energy in discrete shear acceleration}

\begin{table}[tb]
\caption{\label{tab:models}%
The parameter sets for the models shown in Figs. \ref{fig:frac}, \ref{fig:Epeak_simu}, and \ref{fig:bohm}
}
\begin{ruledtabular}
\begin{tabular}{lccccccc}
\textrm{models}&
$R_{\rm jet}$\footnote{In unit of kpc.} & 
$\xi_{c/j}$&
$l_{\rm coh}/R_{\rm coc}$ &
$Z_i$&
$B_{\rm coc}$\footnote{in unit of $\rm \mu$G.} & 
$\beta_{\rm jet}$ & 
$E_{\rm{coh},i}$\footnote{in unit of EeV.}
\\
\colrule
Reference & 0.5 & 10 & 0.03 & 1 & 3 & 0.7 & 0.42 \\
\colrule
A-1 & 0.5 & 10  &0.03 & 26 & 3 & 0.7 & 11 \\
A-2 & 0.5 & 10  & 0.03 & 1 & 15 & 0.7 & 2.1 \\
A-3 & 0.5 & 10  & 0.03 & 1 & 3 & 0.5 & 0.42 \\
B-1 & 0.5 & 4   & 0.03 & 1 & 3 & 0.7 & 0.17 \\
B-2 & 1.5 & 10  & 0.03 & 1 & 3 & 0.7 & 1.2 \\
C-1 & 0.5 & 10  & 0.003 & 1 & 3 & 0.7 & 0.042 \\
C-2 & 0.5 & 100 & 0.003 & 1 & 3 & 0.7 & 0.42 \\
\end{tabular}
\end{ruledtabular}
\end{table}

 \begin{figure}
 \includegraphics[width=\linewidth]{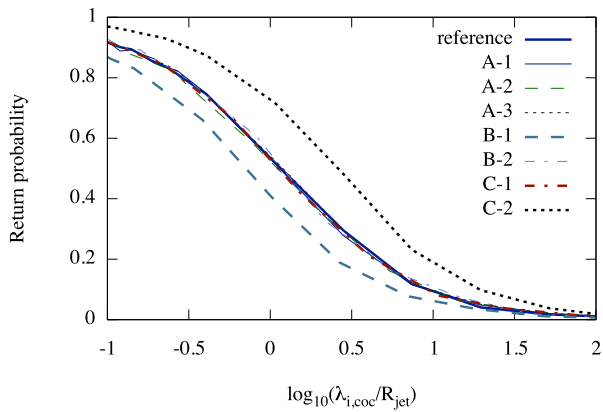}
  \caption{The return probability of the diffusing particles in the cocoon as a function of $\lambda_{i,\rm coc}/R_{\rm jet}$. \label{fig:frac} } 
 \end{figure}

In our cases, the jet confines the CRs more efficiently than the cocoon, i.e. $R_{\rm{jet}}/\lambda_{i,\rm{jet}}>R_{\rm{coc}}/\lambda_{i,\rm{coc}}$, 
{which means that the diffusion in the cocoon determines the maximum energy. If the particle diffusing inside the cocoon returns to the jet, the acceleration cycle continues. Otherwise, the particle escapes from the system.
Fig. \ref{fig:frac} shows the probability of a diffusing particle returning to the jet as a function of $\lambda_{i,\rm coc}/R_{\rm jet}$ for various parameter sets tabulated in Table \ref{tab:models}. The lines completely overlap each other except for B-1 and C-2 that have different values of $\xi_{c/j}$. 
This indicates that the return probability depends only on $\lambda_{i,\rm coc}/R_{\rm jet}$ and $\xi_{c/j}$ under the assumption of $l_{\rm jet}=R_{\rm coc}$. 
Note that the return probability depends on the scattering mean free path not in the jet but in the cocoon. 

When $\lambda_{i,\rm{coc}}\lesssim{R}_{\rm{jet}}$, the majority of the diffusing particles return to the jet after a few random steps. 
This feature does not change regardless of the physical parameters of the jet-cocoon system, as long as $\lambda_{i,\rm coc}< R_{\rm jet}$.
Then, the acceleration time is expressed as 
\begin{equation}
t_{\rm{acc}}=\frac{\Delta{t}}{(\Delta{E}/E)}\sim\zeta_a\frac{\lambda_{i,\rm{coc}}}{c\Gamma_j^2\beta_{\rm jet}^2}, 
\end{equation}
where $\Delta{t}=\zeta_a\lambda_{i,\rm{coc}}/c$ is the typical residence time in the cocoon per cycle
and $\Delta{E}/E\sim\Gamma_{\rm{jet}}^2\beta_{\rm{jet}}^2$ is the mean energy gain per cycle~\cite{Rieger:2004jz}. Here $\zeta_a$ is a correction factor that accounts for the average number of steps over the accelerated particles. 
The residence time in the jet per cycle is much shorter than that in the cocoon, because of the shorter mean free path in the jet.

On the other hand, when $\lambda_{i,\rm{coc}}\gtrsim{R}_{\rm{jet}}$, the majority of the CR particles escape from the cocoon without returning to the jet, as seen  in Fig. \ref{fig:frac}. 
Only the particles that go back to the jet continue to gain energies by the shear. 
Thus, the size of efficient CR acceleration region is limited by the jet size. The effective confinement time in the acceleration region can be represented as 
\begin{equation}
t_{\rm{conf}}=\zeta_c\frac{R_{\rm{jet}}}{c},
\end{equation}
where $\zeta_c=\zeta_c(\xi_{c/j})$ is a geometrical correction factor that takes into account the weak dependence on $\xi_{c/j}$. 
As seen in Fig. \ref{fig:frac}, CRs have a larger chance to return to the jet for a larger $\xi_{c/j}$.
Since the CRs escape through the cocoon, the confinement time itself is not directly related to the mean free path in the jet.

The condition $t_{\rm{acc}}\approx t_{\rm{conf}}$ leads to the maximum energy in the energy spectrum of escaping CRs: 
\begin{equation}
E_{i,\rm max}\approx \zeta e Z_i B_{\rm coc}l_{\rm coc}^{1/2}R_{\rm jet}^{1/2}\Gamma_{\rm jet}\beta_{\rm jet},
\label{eq:Epeak}
\end{equation}
where $\zeta\equiv(\zeta_c/\zeta_a)^{1/2}$ and $\lambda_{i,\rm{coc}}\propto{E^2}$ is used. From the simulation results, we found $\zeta\simeq2.2(\xi_{c/j}/10)^{0.2}$ 
(see Appendix \ref{sec:detail} for the consistency of this estimate and the simulation results),
leading to $E_{i,\rm{max}}\sim1.6Z_i\rm~EeV$ for our reference parameter set (see Fig.~\ref{fig:source_dist}). We confirm this scaling relation for mildly relativistic cases of $\Gamma_{\rm{jet}}\beta_{\rm{jet}}\sim1$~\footnote{Note that for highly relativistic cases, the anisotropy in the momentum distribution would considerably affect $E_{i,\rm{max}}$ in our estimate.}.

The discrete shear acceleration process is one of the Fermi acceleration mechanisms, so the accelerated CRs have a power-law spectrum. 
Almost all the accelerated particles can escape. For $E<E_{i,\rm{max}}$, the escaping CRs show a hard power-law spectrum, $dL_E/dE\propto{E^{-1}-E^0}$ (see Fig.~\ref{fig:source_dist}). It has a spectral break at $E\sim{E_{i,\rm{coh}}}$ due to the change of energy dependence of the mean free path. For $E>E_{i,\rm{max}}$, the spectrum has a cutoff that is slower than the exponential (see Appendix \ref{sec:detail} for the detailed results of Monte Carlo simulations, including the parameter dependence of the spectral shape and cases for the Bohm limit).
Since we consider kiloparsec-scale jets, we can neglect energy losses due to proton synchrotron, hadronuclear, photohadronic, and photodisintegration processes. 

 \begin{figure}
 \includegraphics[width=\linewidth]{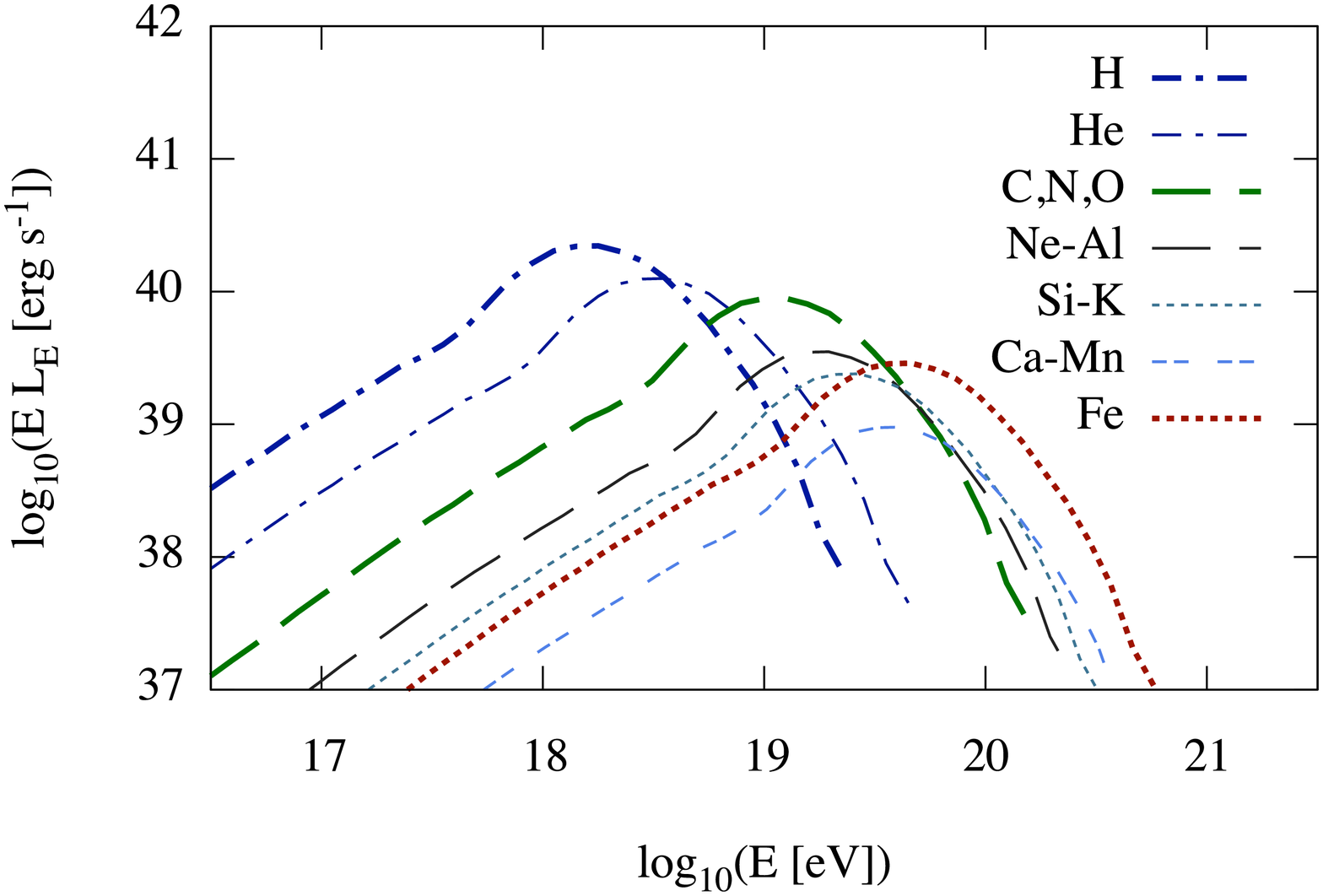}
   \caption{The intrinsic energy spectra of UHECRs produced by shear acceleration with the injection of GCRs. \label{fig:source_dist} } 
 \end{figure}

\subsection{Continuous shear acceleration}\label{sec:continuous}

There is a shear layer between the jet and the cocoon where the jet velocity may change linearly \cite{Aloy:2008zz}. 
This layer affects the spectrum of CRs if the size of shear layer is larger than the Larmor radius or the scattering mean free path of the CRs \cite{Rieger:2006uv}. Here, we make a brief discussion about effects of the shear layer, which may have a crucial influence on the injection process to the discrete shear acceleration (see Section \ref{sec:injection}).

Inside the shear layer, the evolution of distribution function is described by the diffusion equation in momentum space. Adding the escape term and injection term, which are important in our setup, we can write the CR transport equation as \cite{Earl:1988aa,Rieger:2006uv}
\begin{equation}
 \frac{\partial f}{\partial t} = \frac{1}{p^2}\frac{\partial}{\partial p}\left(p^2 D_p \frac{\partial f}{\partial p}\right) - \frac{f}{t_{\rm esc,sl}} + Q_0 \delta(p-p_{\rm inj,sl}),\label{eq:sl}
\end{equation}
where $D_p\approx p^2 \lambda_{i,\rm sl} c (dv_j/dr)^2/15$ is the diffusion coefficient in momentum space ($\lambda_{i,\rm sl}$ is the mean free path and $dv_j/dr$ is the velocity gradient in the shear layer), $t_{\rm esc,sl}=R_{\rm sl}^2/(2\lambda_{\rm sl} c)$ is the escape time from the shear layer ($R_{\rm{sl}}$ is the size of shear layer), $Q_0$ is the injection rate, and $p_{\rm inj,sl}$ is the injection momentum. 
The acceleration time is estimated to be $t_{\rm acc,sl}=p^2/D_p\propto p^{-\delta}$, where we write the mean free path as $\lambda_{i,\rm sl} \approx \lambda_0 (p/p_0)^{\delta}$. This dependence is the same as that of $t_{\rm esc}$, which means that the acceleration time is shorter for higher energy for $\delta>0$ \cite{Rieger:2006uv}.
Assuming a power-law distribution function $dN/dE=4\pi p^2f \propto p^{-s_{\rm sl}}$, we can obtain the steady state solution as 
\begin{equation}
  s_{\rm sl} = \left\{ \begin{array}{ll}
		\frac{\delta-1}{2} - q_{\rm sl} & (p<p_{\rm inj,sl}) \\
			\frac{\delta-1}{2} + q_{\rm sl}   & (p>p_{\rm inj,sl})
  \end{array} \right.,
\end{equation}
\begin{equation}
 q_{\rm sl}=\sqrt{\frac{(\delta+3)^2}{4}+\frac{t_{\rm acc,sl}}{t_{\rm esc,sl}}}.
\end{equation}
We confirm this power-law solution by numerically solving Eq. (\ref{eq:sl}). The spectrum of escaping particles is written as $dL_E/dE\approx (dN/dE)/t_{\rm esc,sl}\propto E^{-s_{\rm esc,sl}}$, where $s_{\rm esc,sl}=\mp q_{\rm sl}-(1+\delta)/2$. Considering the linear velocity gradient, $(dv_j/dr)\approx c\beta_{\rm jet}/R_{\rm sl}$, we obtain $t_{\rm acc,sl}/t_{\rm esc,sl}=30 c^2/(R_{\rm sl}^2 (dv_j/dr)^2 )\approx 30 \beta_{\rm jet}^{-2}$. Then, the index of the escape spectrum is $s_{\rm esc,sl}\sim 7.3~(-8.7)$ for $p>p_{\rm inj}$ ($p<p_{\rm inj}$). This spectrum is so steep that it cannot match the observed UHECR spectrum. Most of the injected particles escape from the shear layer before being accelerated to higher energies. 
In other words, only few low-energy GCRs that are injected to the continuous shear acceleration can reach the injection energy, above which the discrete shear acceleration operates (see Section \ref{sec:injection}).
Thus, the low-energy GCRs are unlikely to be accelerated to UHECRs.
Here, we assume that the particles are injected at the center of the shear layer for simplicity. In reality, the particles are injected at the edge of the shear boundary. Although this could affect the spectral shape, it is unlikely that the injection position drastically changes the acceleration efficiency.
More detailed discussions for the continuous shear acceleration are beyond the scope of this work, and remains as a future work.

\section{Recycling Galactic CRs as UHECRs}\label{sec:recycling}

\subsection{Injection rate and composition ratio}\label{sec:injection}

In our shear reacceleration scenario, we have shown that the spectrum of escaping CRs is generically hard, and $E_{i,\rm{max}}$ is determined by the five parameters ($\beta_{\rm{jet}}$, $R_{\rm{jet}}$, $\xi_{c/j}$, $l_{\rm{coh}}$, $B_{\rm{coc}}$). Next, we estimate the UHECR luminosity and their composition ratio.

CR densities in radio galaxies are highly uncertain. Here, we assume that the proton CR densities are comparable to that in our Galaxy. 
While the star-formation rate of elliptical galaxies may be lower than that of star-forming galaxies by a factor of 3--10~\cite{Salmi:2012aa,Martig:2012vc}, this uncertainty is easily absorbed by uncertainties in the other parameters. 
The GCR density inside the CR halo of $H_h\sim5$~kpc \cite{Strong:2007nh} can be expressed as 
\begin{equation}
 n_{i,d}=K_i\left(\frac{E_{i,\rm{inj}}}{{\rm{TeV}}}\right)^{-\alpha_i+1}\exp\left(-\frac{E_{i,\rm{inj}}}{Z_i{\rm~PeV}}\right).
\end{equation}
Here, CR species are grouped as $i=$ H, He, C--O, Ne--Al, Si--K, Ca--Mn, Fe. Their effective charge $Z_i$ and atomic mass $A_i$ are $Z_i=$ 1, 2, 7, 11, 15, 23, 26 and $A_i=$ 1, 4, 14, 23, 30, 49, 56, respectively. 
We use the observed values at $E\sim1$ TeV for the normalization of each component: $K_{\rm{H}}=3.6\times10^{-15}\rm~cm^{-3}$ and $K_i/K_{\rm{H}}\simeq$ 1, 0.65, 0.33, 0.17, 0.14 0.072, 0.23~\cite{WiebelSooth:1997yc,Hoerandel:2002yg}. 
In the galactic disk, the proton has softer index than the others~\cite{WiebelSooth:1997yc,Hoerandel:2002yg,Caprioli:2010ne,Yoon:2011aa}, $\alpha_{\rm{H}}\simeq2.7$ and $\alpha_{i\neq\rm{H}}\simeq2.6$~\footnote{This is conservative and would be more appropriate for older galaxies. Harder spectra assumed in~\cite{Caprioli:2015zka} relaxes the energetics requirement}. 
In addition, we increase the abundance of nuclei heavier than He by factor of 3 from the value above because most of radio galaxies have more metals than the Galaxy due to their past star formation activities~\cite{Henry:1999bn,Tang:2009aa}.

The number of swept-up particles of species $i$ by the time when $l_{\rm{jet}}=H_h$ is simply given by $2\pi R_{\rm{coc}}^2H_hn_{i,d}$, and we assume that only the fraction, $R_{\rm{jet}}^2/R_{\rm{coc}}^2$, is injected into shear acceleration. 
Thus, the time-integrated number of injected GCRs are written as $N_{i,\rm inj}\approx2\pi R_{\rm jet}^2H_hn_{i,d}$. The swept-up particles of $\lambda_{i,\rm sl}<R_{\rm{sl}}$ are accelerated by the continuous shear that is ineffective to produce high-energy CRs (see Section \ref{sec:continuous}).
Only the particles of $\lambda_{i,\rm sl}>R_{\rm{sl}}$ can be injected to the discrete shear acceleration process. 
Setting $\lambda_{i,\rm sl}=R_{\rm sl}$, the injection energy is given by $E_{i,\rm{inj}}\approx E_{\rm coh}(R_{\rm{sl}}/l_{\rm{coh}})^3\sim15Z_i\rm~TeV$. 
Here, we use $\lambda_{i,\rm sl}\sim\lambda_{i,\rm coc}$ and $R_{\rm{sl}}\sim0.01R_{\rm{jet}}\sim5$~pc.
The injected CRs are accelerated until the adiabatic cooling is effective, $t_{\rm{ad}}\approx{R}_{\rm{coc}}/v_{\rm{exp}}\sim1.6$~Myr (where $v_{\rm{exp}}\sim3000\rm~km~s^{-1}$~\cite{Bordas2011:aa}). The time-averaged injection rate of GCRs of species $i$ to shear acceleration is estimated to be 
\begin{equation}
 \dot{N}_{i,\rm{inj}}\approx\frac{N_{i,\rm inj}}{t_{\rm ad}}\approx\frac{2\pi{R}_{\rm{jet}}^2H_h n_{i,d}}{t_{\rm{ad}}}.
\end{equation}

Renormalizing the simulation input by the injection rate, we obtain the differential luminosity of UHECRs, $L_{\rm{UHECR}}$. The CR luminosity density at ${10}^{19.5}$~eV is $0.6\times{10}^{44}{\rm~erg~Mpc^{-3}~yr}^{-1}$ (e.g.,~\cite{Murase:2008sa}), and the number density of FR Is is roughly $\sim{10}^{-5}-{10}^{-4}\rm~Mpc^{-3}$~\cite{Padovani:2011wj,Prescott:2016aa}. Thus, $L_{\rm{UHECR}}\sim2\times10^{40}-2\times{10}^{41}~\rm~erg~s^{-1}$ is required. Our model can satisfy this requirement, as shown in Fig.~\ref{fig:source_dist}. 
Also, our model can avoid anisotropy constraints at $E\sim10$~EeV~\cite{AbuZayyad:2012hv} owing to the high source number density with the heavy composition.
The relative abundance ratio at the same rigidity is estimated to be $(f_{\rm{H}},~f_{\rm{He}},~f_{\rm{C\mathchar`-O}},~f_{\rm{Ne\mathchar`-Al}},~f_{\rm{Si\mathchar`-K}},~f_{\rm{Ca\mathchar`-Mn}},~f_{\rm{Fe}})=$ (0.73, 0.21, 0.042, 0.011, 0.0053, 0.0014, 0.0037). Note that we cannot freely change the abundance ratio among heavy nuclei as well as the intrinsic spectral index, because they are determined by the shear acceleration mechanism and observed abundance of Galactic CRs.

\subsection{Comparison with observations}

We calculate the propagation of the UHECRs from the sources to the Earth using CRPropa 3~\cite{Armengaud:2006fx,Batista:2016yrx}. The code includes the photomeson production, the photodisintegration, and the electron-positron pair production through the cosmic microwave background and extragalactic background light (EBL).  The nuclear decay process is also included. We use the EBL model of~\cite{Finke:2009xi}, and assume that all FR Is produce the UHECRs shown in Fig.~\ref{fig:source_dist} for simplicity. 
The luminosity density of bright AGNs positively evolves with redshift~\cite{Ajello:2013lka,Ueda:2014tma}, while that of low-luminosity AGNs may have a weaker redshift evolution~\cite{Ajello:2013lka,Padovani:2011wj}. In this work, we assume no redshift evolution but stronger evolution models can also fit the data. 

We show the spectrum of the UHECRs at the Earth in the upper panel of Fig.~\ref{fig:propa}. The intermediate and heavy nuclei decrease while protons increase during the propagation process due to the photodisintegration. The cutoff at $E\gtrsim100$ EeV is produced due to the maximum energy of the shear acceleration at the source, which is consistent with the PAO data. We need an additional component to fit the spectrum at $E\sim$ EeV (e.g.,~\cite{Katz:2008xx,Aloisio:2013hya,Thoudam:2016syr}). The middle panel and the lower panel show the mean depth of the shower maximum, $\langle{X}_{\rm{max}}\rangle$, and variance of the shower depth, $\sigma(X_{\rm{max}})$, respectively. These values are calculated using $X_{\rm{max}}$ probability distribution parametrized by~\cite{DeDomenico:2013wwa}. Within systematic errors, our model reasonably explains the observed feature of the chemical composition that changes from light to heavy as CR energy increases, without tuning the abundance ratio by hand.

\begin{figure}
\includegraphics[width=\linewidth]{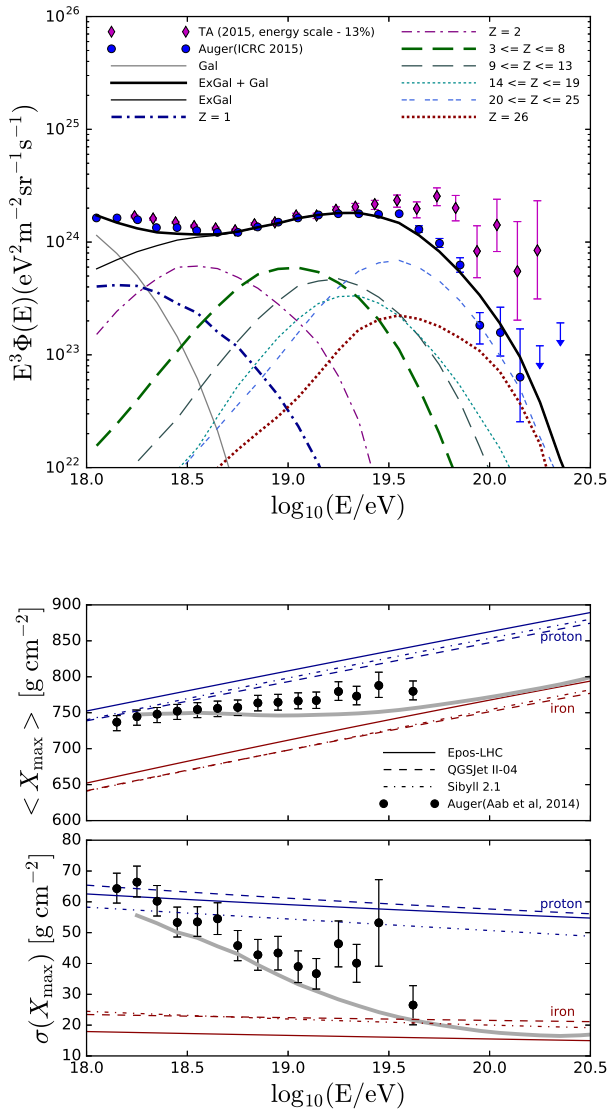}
\caption{The observed spectrum (upper panel), $\langle{X_{\rm{max}}}\rangle$ (middle panel), and $\sigma(X_{\rm{max}})$ (lower panel) of the UHECRs at the Earth. The data of PAO and TA are taken from~\cite{Aab:2014kda,Aab:2015bza,Fukushima:2015bza}.\label{fig:propa} } 
\end{figure}

\section{Summary and discussion}\label{sec:summary}

We have shown that the shear acceleration by black-hole jets provides a promising mechanism of UHECR production. Based on the setup for the jet-cocoon system that is ubiquitous in radio galaxies, we have performed detailed numerical simulations of UHECR acceleration, escape, and propagation in intergalactic space. The radio galaxies can accelerate protons up to a few EeV and irons up to 100 EeV, whose spectra are intrinsically hard as required by the PAO data. TeV--PeV CRs in a galactic halo are injected to the shear acceleration, leading to the enhanced metal abundance suggested by the $\langle{X_{\rm{max}}}\rangle$  and $\sigma(X_{\rm{max}})$ data.

We stress that the spectrum and composition are essentially determined by theoretical calculations and observations of Galactic CRs, respectively. Although the calculation of propagation is slightly affected by the redshift evolution of the sources and the EBL model~\cite{Aab:2016zth}, this cannot change our conclusion. It is possible to alter the source spectral index by superposing contributions from radio galaxies that have different $E_{i,\rm{max}}$. While more luminous radio galaxies could accelerate UHECRs to higher energies, $L_{\rm{UHECR}}$ is independent of the jet luminosity if all the radio galaxies have the same size of the halos. Then, fainter radio galaxies such as FR Is may give the most important contribution to the observed UHECR flux. On the other hand, the source parameters, such as $l_{\rm{coh}}$, $\beta_{\rm{jet}}$, and $R_{\rm{sl}}$, are uncertain. Phenomenologically, all the uncertainties are absorbed by treating $E_{p,\rm{max}}$ and $L_{\rm{UHECR}}$ as free parameters.  
The source models with similar values of $E_{p,\rm{max}}$ give the similar shape of the spectra at the Earth, $\langle{X}_{\rm{max}}\rangle$, and $\sigma(X_{\rm{max}})$. According to observations and simulations of the jet propagation~\cite{Laing:2002aa,Aloy:2008zz}, $R_{\rm{sl}}/R_{\rm{jet}}\sim0.1$ and $\beta_{\rm{jet}}\sim0.9$ are also possible, where we would need smaller $B_{\rm{coc}}$ and larger $l_{\rm{coh}}$ to obtain the required $E_{p,\rm{max}}$ and $L_{\rm{UHECR}}$.

We have considered shear acceleration in large-scale jets, which is different from the scenario by~\cite{Murase:2011cy} for UHECR acceleration in blazar jets. Our model is also different from~\cite{Caprioli:2015zka}, which relies on the first encounter boost in the relativistic jet of $\Gamma\sim30$~\cite{Gallant:1998uq}, whereas both consider the injection of Galactic CRs. While such jets could exist in sub-parsec scales as suggested in blazars or even kiloparsec scales for the most powerful FR II galaxies, jets of FR Is are significantly decelerated in such large scales, and mildly relativistic jets are considered in this work~\cite{Bowman1996,Laing:2002aa,Canvin2005:aa}. 

FR Is and their blazar counterparts, BL Lac objects, are observed at different wavelengths from radio to gamma-rays. The charged particles that emit the observed electromagnetic signals are likely to be produced at different locations in the shear layer, e.g., by internal shocks~\cite{Rees:1978aa} or turbulence \cite{Kakuwa:2015fda}. In the leptonic scenario, the electrons are difficult to get accelerated solely by the discrete shear acceleration mechanism, since their typical energy is lower than $E_{i,\rm{inj}}$~\cite{Inoue:1996vv}.

Our model is consistent with the convergence picture of UHECRs, neutrinos, and gamma rays~\cite{Murase:2016gly,Fang:2017zjf}, in which all three messengers are explained simultaneously. In the galaxy cluster and group model, UHECRs can be provided by AGNs~\cite{Fang:2017zjf}. CRs that do not reach ultrahigh energies can be accelerated by the AGN jet without the shear reacceleration, and the CR spectrum can be effectively extended to ultrahigh energies with a hard spectrum via the shear acceleration mechanism.    
Also, the corresponding cosmogenic neutrino flux is expected to be $\sim{10}^{-10}{\rm~GeV~cm^{-2}~s^{-1}~sr^{-1}}$. Gamma rays and neutrinos associated with large scale jets may not be easy to detect due to long energy-loss time scales (cf. \cite{Ostrowski:1999py}). Whereas electrons may be difficult to be injected into the shear acceleration process, it is important to study indirect signatures through radio and/or X-ray observations~\cite{Stawarz:2002uh,Laing:2002aa,Stawarz:2005tq} for testing our model.


\medskip
\begin{acknowledgments}
We acknowledge Damiano Caprioli, Martin Lemoine, Miguel Mostafa, Michael Ostrowski, and G\"{u}nter Sigl for useful discussion. 
The work is supported by Alfred P. Sloan Foundation, NSF Grant No. PHY-1620777 (K.M.), NASA NNX13AH50G, and the IGC post-doctoral fellowship program (S.S.K.). B.T.Z. is supported by China Scholarship Council (CSC) to conduct research at Penn State University.
Monte Carlo simulations in this work were carried out on Cray XC30 at Center for Computational Astrophysics, National Astronomical Observatory of Japan and on the cluster Draco in Tohoku University.
\end{acknowledgments}

\appendix
\section{Details of Monte Carlo simulations}\label{sec:detail}

In this appendix, we describe the results of Monte Carlo simulations, focused on the situation that $R_{\rm coc}/\lambda_{i,\rm coc}< R_{\rm jet}/\lambda_{i,\rm jet}$ and $l_{\rm coh}< R_{\rm jet}$.

\subsection{Parameter dependence}

 \begin{figure}
 \includegraphics[width=\linewidth]{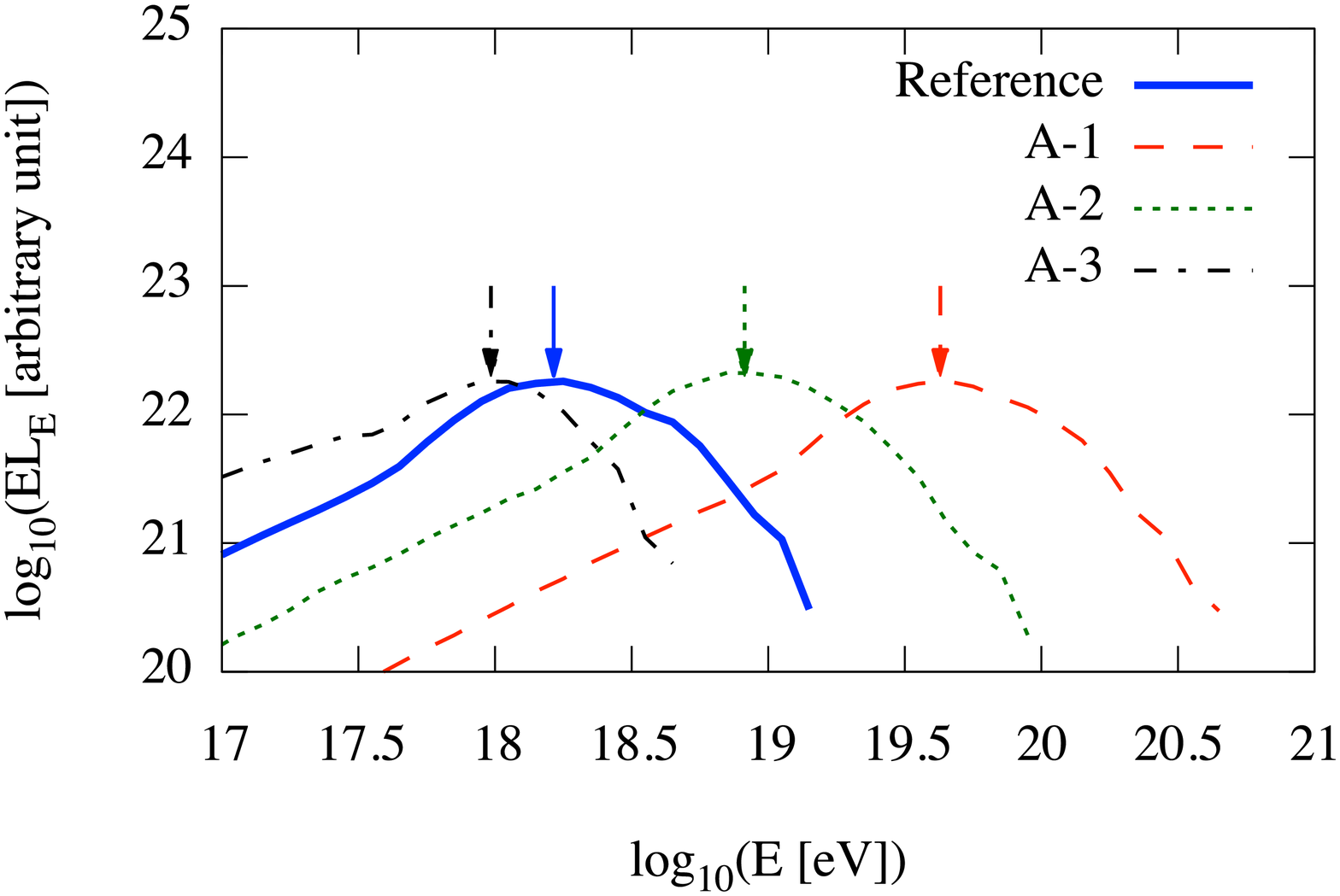}
 \includegraphics[width=\linewidth]{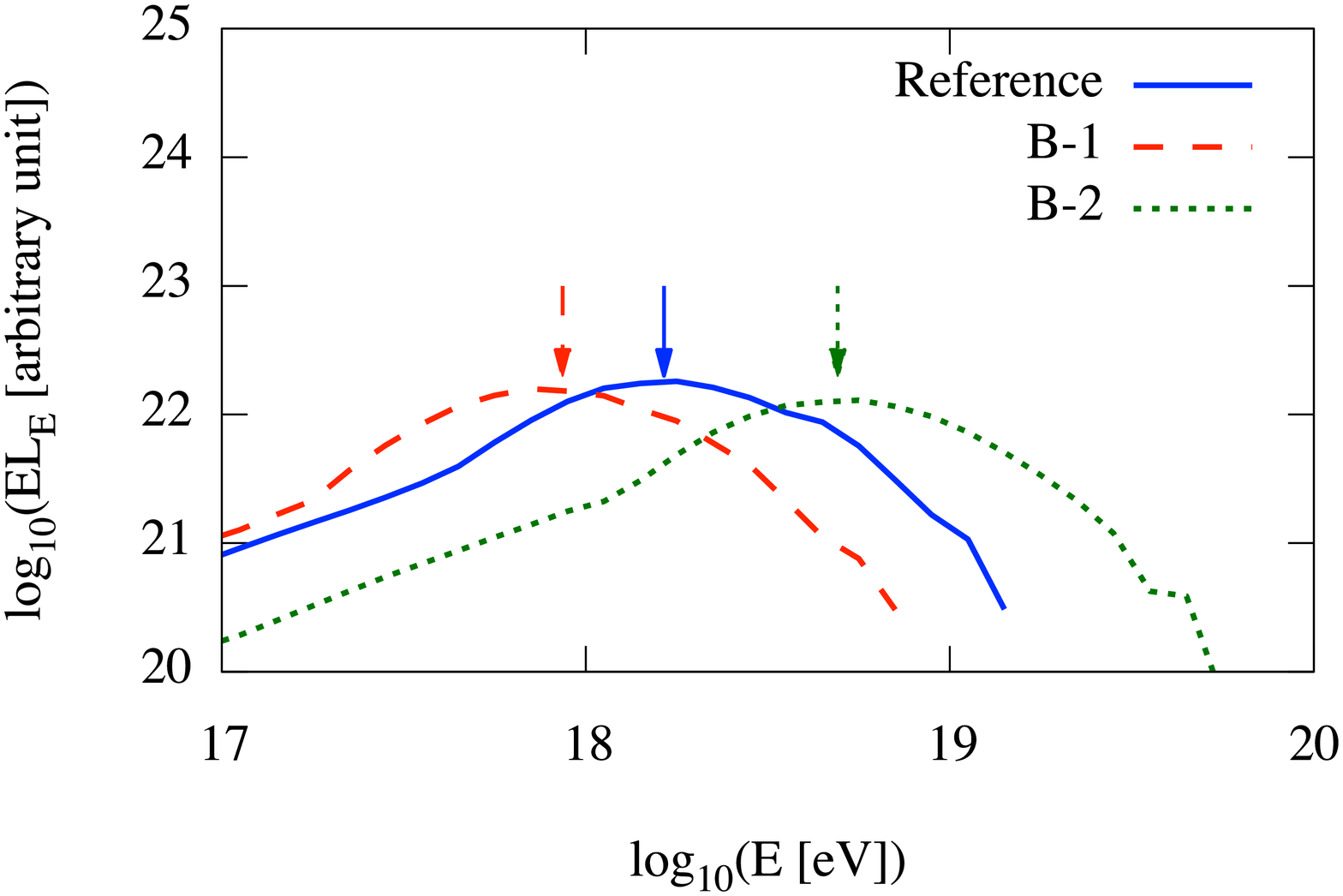}
 \includegraphics[width=\linewidth]{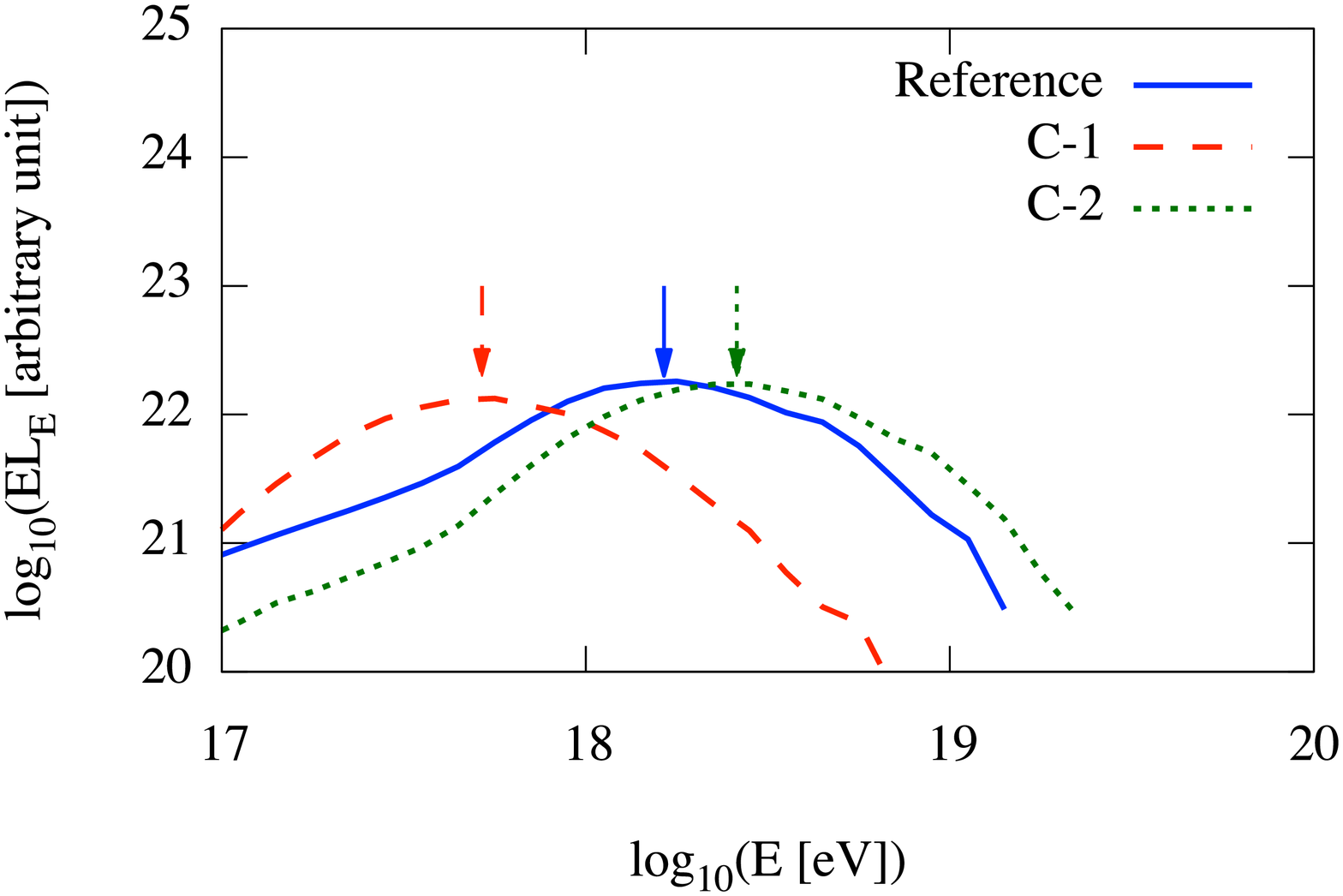}
  \caption{The results of the escape spectra with various parameter sets. The lines show the escape spectra, and arrows show the estimated peak energy with $\zeta=2.2(\xi_{c/j}/10)^{0.2}$.
\label{fig:Epeak_simu} } 
 \end{figure}

As discussed in the main text, we obtain the maximum energy by setting $t_{\rm acc}=t_{\rm esc}$, which results in
\begin{equation}
E_{i,\rm max}\approx \zeta e Z_i B_{\rm coc}l_{\rm coh}^{1-1/\delta}R_{\rm jet}^{1/\delta} \Gamma_{\rm jet}^{2/\delta} \beta_{\rm jet}^{2/\delta}.
\label{eq:Epeak_sm}
\end{equation}
We perform Monte Carlo simulations with various parameter sets tabulated in Table \ref{tab:models} to see the values of $\zeta$. The results are shown in Fig.~\ref{fig:Epeak_simu}, where the lines represent the escape spectra and the corresponding arrows show the peak energy estimated by Eq. (\ref{eq:Epeak_sm}) with $\delta=2$ and $\zeta\simeq2.2(\xi_{c/j}/10)^{0.2}$. We can see that the simulation results agree with the estimates well.

According to our simulation results, the spectral shape is not sensitive to the parameters for $E\gtrsim E_{\rm coh}$, as seen in Fig.~\ref{fig:Epeak_simu}. We try to fit the spectral shape there using a combination of a power law growth and a cutoff. We consider $EL_E\propto E^a \exp(-(E/E_0)^b)$, and find that $a\sim$ 5--9 and $b\sim$ 0.1--0.3 for the parameter range that we explored. Note that the fitting requires $a>1$, $b<1$, and $E_0\ll E_{i,\rm max}$ because of slower cutoff than the exponential.

\subsection{Bohm diffusion model}

We also perform Monte Carlo simulations using the Bohm limit in the cocoon, $\lambda_{i,\rm coc}=E/(Z_i e B_{\rm coc})$ for $E < E_{i,\rm coh}$ and $\lambda_{i,\rm{coc}}=(E/E_{i,\rm{coh}})^2l_{\rm{coh}}$ for $E > E_{i,\rm coh}$. Figure \ref{fig:bohm} shows the escape spectra for the cases with the Bohm limit. For these cases, the maximum energy is represented by Eq. (\ref{eq:Epeak_sm}), while the spectra for $E<E_{i,\rm coh}$ is harder than those with the Kolmogorov turbulence. This difference arises from the difference of energy dependence of the mean free path. For the particles of $E<E_{i,\rm coh}$, the mean free path for the Bohm limit is shorter than that for the Kolmogorov turbulence. The shorter mean free path leads to the higher return probability, which results in the harder escape spectrum for the Bohm limit cases. 

The shorter mean free path in the cocoon also increases the value of $E_{i,\rm inj}$, leading to the lower $\dot N_{i,\rm inj}$ and $L_{\rm UHECR}$. To obtain the required $L_{\rm UHECR}$ and $E_{p,\rm max}$, we would need lower $B_{\rm coc}$ and higher $\beta_{\rm jet}$.

 \begin{figure}
 \includegraphics[width=\linewidth]{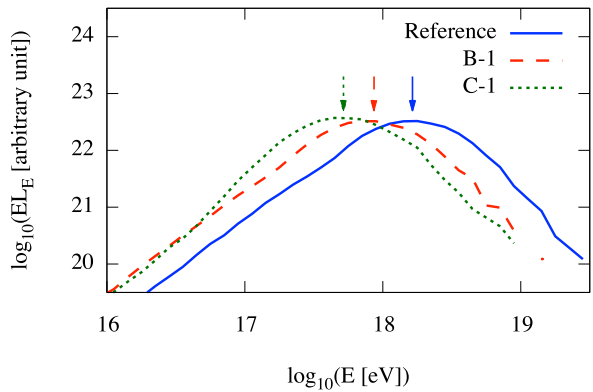}
  \caption{The results of the escape spectra for the cases with the Bohm limit. The lines show the escape spectra, and arrows show the estimated peak energy.
\label{fig:bohm} } 
 \end{figure}

\section{Application to small-scale jets of radio-quiet AGNs}\label{sec:radio-quiet}

 \begin{figure}[tb]
 \includegraphics[width=\linewidth]{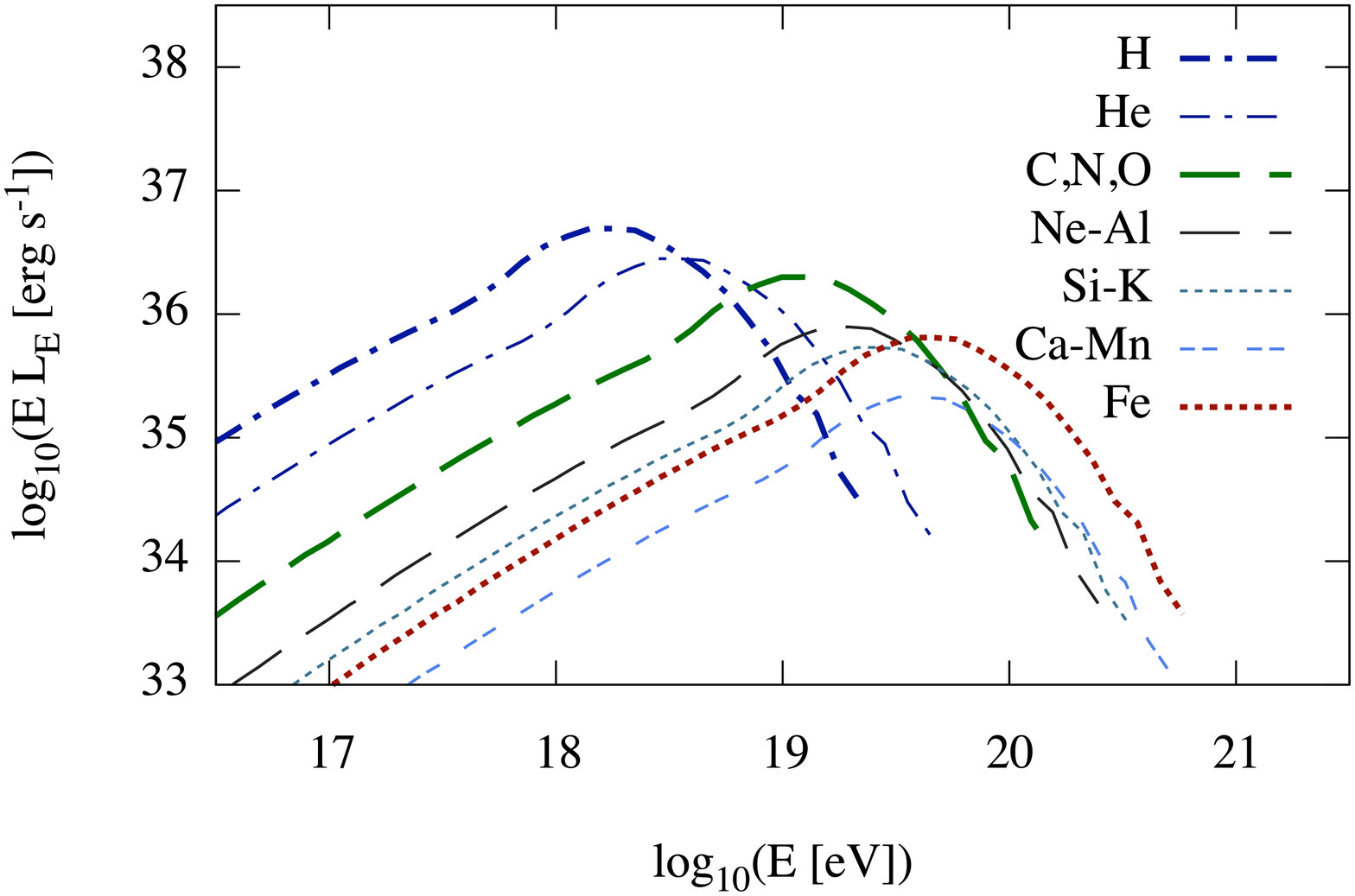}
   \caption{The source spectrum for the radio-quiet AGN model. \label{fig:seyfert_source} } 
 \end{figure}

Our model can be applied to the small-scale jets in radio-quiet AGNs (e.g.,~\cite{Mundell:2002hy,Ho:2008rf}),  which are energetically possible to create UHE CRs \cite{Peer:2009vnw}. The jet size there is much smaller than that in the radio galaxy. We use $R_{\rm jet}\sim 10$ pc, $l_{\rm jet}\sim 200$ pc, and $R_{\rm coc}\sim 100$ pc. The magnetic field can be stronger than that in the radio galaxy owing to its smaller size. We use $B_{\rm coc}\sim 160\rm~\mu$G. This jet is embedded in galactic center, so the CR density and metalicity can be enhanced, compared to those around the Earth. We use 20 times higher CR density~\cite{Abramowski:2016mir} and 2 times higher metallicity~\cite{Rudolph:2006abc} than those in the Galaxy described in the main text. At the center of the radio-quiet AGNs, the outflows of velocity $\sim$100--1000 km s$^{-1}$ are observed~\cite{Steffen:1997tw}, and we use $v_{\rm exp}\sim 1000$ km s$^{-1}$.  We set the other parameters to be the same as those for the radio gaalxy; $\beta_{\rm jet}\sim 0.7$,  $R_{\rm sl}\sim 0.01R_{\rm coc}$, and $l_{\rm coh}\sim 0.03R_{\rm coc}$. 

Using above parameters, we obtain the source spectrum as shown in Fig.~\ref{fig:seyfert_source}. The radio-quiet AGNs can accelerate protons up to a few EeV and irons up to several tens of EeV. The number density of radio-quiet AGNs is around $10^{-3}\rm~Mpc^{-3}$ \cite{Ho:2008rf}, so the required differential luminosity per source is $3\times10^{39}\rm~erg~s^{-1}$. However, we find that the radio-quiet AGN model does not reach the required luminosity. It is difficult for the radio-quiet AGN model to achieve both the required values of $E_{p,\rm max}$ and $L_{\rm UHECR}$.


\bibliography{shear}
\bibliographystyle{apsrev}
\end{document}